\documentclass[preprint,review,12pt]{elsarticle}

\ifpreprint
\fi
\usepackage{lmodern}

\usepackage{graphics}
\usepackage{graphicx}

\usepackage{booktabs}
\usepackage{multirow}
\usepackage{array}
\usepackage{amsmath}
\usepackage{amsthm}
\usepackage{verbatim}
\usepackage{relsize}
\usepackage[dvipsnames]{xcolor}
\usepackage{array}
\usepackage{bm}

\usepackage{caption}

\usepackage{algorithm}
\usepackage{algpseudocode}

\usepackage{siunitx}
\usepackage{natbib}
\usepackage{hyperref}
\newcommand{\doi}[1]{\textsc{doi}: \href{http://dx.doi.org/#1}{\nolinkurl{#1}}}

\definecolor{comment}{rgb}{0.0, 0.5, 0.0}

\usepackage{hyperref}

\hypersetup{
    colorlinks=true,
    linkcolor=blue,
    filecolor=magenta,
    urlcolor=cyan,
}
\usepackage{booktabs}
\usepackage{xcolor}
\usepackage{color}

\definecolor{mygreen}{rgb}{0,0.6,0}
\definecolor{mygray}{rgb}{0.5,0.5,0.5}
\definecolor{mymauve}{rgb}{0.58,0,0.82}

\usepackage{pdfcomment}

\usepackage{array}
\usepackage{mathtools}
\usepackage{wasysym}
\usepackage{verbatim}
\usepackage{relsize}
\usepackage{multirow}
\usepackage{multicol}

\usepackage{listings}
\newcommand{\bs}[1]{\boldsymbol{#1}}

\usepackage[bitstream-charter]{mathdesign}
\usepackage[T1]{fontenc}
\usepackage[utf8]{inputenc} 

\usepackage[english]{babel}
\usepackage[normalem]{ulem}

\DeclareMathAlphabet{\pazocal}{OMS}{zplm}{m}{n}

\def \A {\pazocal{A}}

\usepackage{lscape}		
\usepackage{fancyvrb}

\usepackage[mathscr]{euscript}

\DeclareMathOperator*{\argminA}{arg\,min} 

\usepackage{textcomp}

\def \balpha {{\bm{\alpha}}}
\def \bbeta  {{\bm{\beta}}}

\def \bPsi {{\boldsymbol\Psi}}
\def \xxi  {{\boldsymbol{\xi}}}

\def \X  {\bm{X}} 
\def \x  {\bm{x}} 

\newcommand{\ns} {{\ensuremath{n_{\mathrm{sim}}}}}
\newcommand{\der}{\mathop{}\!{d}}
\def \PCE {\textsf{PCE}}

\usepackage{bigints}

\usepackage[switch]{lineno}

  \usepackage{lineno}

\journal{}

\begin{document}

\begin{frontmatter}



\title{On Fractional Moment Estimation from Polynomial Chaos Expansion}


\author{Luk{\'a}{\v s} Nov{\'a}k\corref{cor1}}  \ead{novak.l@fce.vutbr.cz}  \cortext[cor1]{Corresponding author}

\address{Brno University of Technology, Brno, Czech Republic}

\author{Marcos Valdebenito}              \ead{marcos.valdebenito@tu-dortmund.de} 
\author{Matthias Faes}                     \ead{matthias.faes@tu-dortmund.de}
\address{TU Dortmund University, Dortmund, Germany}

\begin{abstract}

Fractional statistical moments are utilized for various tasks of uncertainty quantification, including the estimation of probability distributions. 
However, an estimation of fractional statistical moments of costly mathematical models by statistical sampling is challenging since it is typically not possible to create a large experimental design due to limitations in computing capacity. 
This paper presents a novel approach for the analytical estimation of fractional moments, directly from polynomial chaos expansions. 
Specifically, the first four statistical moments obtained from the deterministic PCE coefficients are used for an estimation of arbitrary fractional moments via H\"{o}lder's inequality.   
The proposed approach is utilized for an estimation of statistical moments and probability distributions in three numerical examples of increasing complexity. 
Obtained results show that the proposed approach achieves a superior performance  in estimating the distribution of the response, in comparison to a standard Latin hypercube sampling in the presented examples.

\end{abstract}

\begin{highlights}
\item Fractional moments are estimated analytically from polynomial chaos expansion.
\item The proposed method is based on well-known H\"{o}lder's inequality.
\item Fast and accurate estimates obtained for low-size experimental designs.
\item Estimated fractional moments allow accurate estimation of probability densities.
\end{highlights}


\end{frontmatter}

\section{Introduction}

Mathematical models of the response $\mathbf{Y}$ of physical systems can be generally represented by functions $\pazocal{M}$ of input vectors $\mathbf{X}$, which provide a mapping $\pazocal{M}: \mathbb{R}^{n_x}\mapsto\mathbb{R}^{n_y},~\mathbf{X}\to\mathbf{Y}$. 
Input variables $\mathbf{X}$ representing physical quantities (e.g. material parameters, geometrical properties, applied loads) may be affected by a certain level of uncertainty. 
Therefore, it is necessary to propagate the uncertainty associated with input variables described by specific probability distributions through the mathematical model in order to obtain realistic results predicting the model's response $\mathbf{Y}$ and its uncertainty.
The task of the analyst is then in this case to perform uncertainty quantification (UQ) of the model response $\mathbf{Y}$, also called quantity of interest (QoI). 
In the simplest case, UQ can be based on pseudo-random sampling of the input random vector and performing corresponding repetitive evaluations of the deterministic model~$\pazocal{M}(\mathbf{X})$. 
Obtained set of results can be further statistically processed to get statistical moments, and ultimately probability distribution of the QoI. 
The estimation of probability distribution function (PDF) or cumulative distribution function (CDF) from given set of statistical moments is not a trivial task and thus, there are various specialized methods for this purpose. 
On the one hand, it is possible to assume a known specific family of probability distributions and fit a PDF to given data. 
Although this is a simple approach requiring typically low number of statistical samples, an assumption of a probability distribution may significantly affects obtained results in further steps of UQ and/or reliability analysis.
On top, it inherently introduces a measure of subjectivity into the analysis, which might not be warranted for critical applications.
On the other hand, one can construct an arbitrary distribution function numerically, e.g. by kernel density estimation \cite{Davis2011}. 
Numerically constructed distributions offer high versatility, though it is typically necessary to optimize hyper-parameters associated with them.  
Usually, an optimal balance between numerical efficiency and flexibility is offered by artificial distribution models parameterized by statistical moments. 
Classic representatives of parameterized distributions are Gram-Charlier expansion or Edgeworth series expansion \cite{GramConditions} based on perturbation of a Gaussian probability distribution. 
More recent developments offer for example three-parameter lognormal distribution \cite{ZHAO200147},  Hermite model \cite{10.1115/1.1288028}, cubic normal distribution \cite{ZHAO20181}, generalized lambda distribution \cite{ZHU2021107815} or distribution functions that are parametrized by a higher number of moments  or even fractional moments \cite{DING2023109775}. 
In this paper, we adopt the M-EIGD-LESND function, which is in essence a \underline{m}ixture of an \underline{e}xtended \underline{i}nverse \underline{G}aussian \underline{d}istribution and a \underline{l}og \underline{e}xtended \underline{s}kew-\underline{n}ormal
\underline{d}istribution (note that underlined letters explain the acronym M-EIGD-LESND). 
It is fully characterised by a set of eight parameters, making it highly flexible, and hence, powerful to fit any type of distribution on $\mathbf{Y}$.

Despite the flexibility of the approaches described above, they require a significant number of samples to allow them to represent the \textit{real} distribution of $\mathbf{Y}$ accurately.
Unfortunately, a combination of sampling-based methods with costly mathematical models is highly time-consuming or even not feasible in industrial applications and surrogate models are often utilized as computationally efficient approximations of the original mathematical model.
There are various types of surrogate models (eg. artificial neural networks, Kriging, support vector machines), with polynomial chaos expansion (\PCE) being a very popular method for UQ. 
\PCE{} was originally proposed by Norbert Wiener \cite{Wiener_PCE} and further investigated in the context of engineering problems by many researchers, e.g.~\cite{SUDRET,Ghanem_spectral}, and it provides an efficient tool for estimation of statistical moments and sensitivity indices. 
Especially the \PCE{} in its non-intrusive form (spectral projection and regression) possesses significant potential for industrial applications, since it offers a~convenient way to perform advanced probabilistic analysis of any black-box model without any modifications of existing numerical solvers. 
In practice, it becomes often necessary to employ sparse \PCE{}s that yield efficient solutions for real-world physical systems. 
Regression-based non-intrusive \PCE{} \cite{BLATMANLARS} offers large variety of solvers \cite{LuthenReview}, sampling schemes \cite{Conover:LHS:75,VorEli:Technometrics:20,CoherenceOptPCE,InducedSampling} and adaptive algorithms \cite{PCESoptimSeq,SequentialPCEThapa,ZhouSequential,NOVAK2021114105} leading to a large variety of methods. 
Once a \PCE{} is available for a given mathematical model, the constructed explicit function can be exploited to obtain additional information about that model.
This information includes integer statistical moments\cite{SUDRET}, probability distribution of QoI or sensitivity indices \cite{ZHU2021107815, NOVAK2022106808}, which can be calculated without additional evaluations of the underlying numerical model $\pazocal{M}$, which is especially beneficial in industrial applications \cite{fibPCE,CRESTAUX20091161}. 

Despite many recent advances in the field of \PCE{}, the challenge of estimating the distribution of $\mathbf{Y}$, especially in its tails, is still open.
A particularly interesting route to estimate this distribution is through the estimation of fractional moments of $\mathbf{Y}$, since it can be shown that they carry information about an infinite number of integer moments~\cite{DING2023109775}.
This, in its turn, could potentially allow for a more accurate estimation of the distribution of $\mathbb{Y}$ following a moment matching procedure, see e.g.  \cite{Zhang_2022h}.
This paper is therefore focused on estimation of fractional moments directly from \PCE{}, and their further utilization for an approximation of probability distribution of QoI by adopting a recently proposed distribution parameterized by fractional moments.
Section~\ref{PCE section} gives a brief introduction of the main mathematical concepts concerning \PCE{} that are required to understand the developments later in the paper.
Secion~\ref{sec:FractionalMoments} introduces the concept of fractional moments, and how they can be estimated analytically from a trained \PCE.
Section~\ref{sec:NumericalExamples} illustrates the developments and their efficacy using three numerical examples, ranging from an analytical function, over a finite element model of a plate in bending, to a dynamically loaded mass-spring system.
Section~\ref{sec:Conclusions} lists the conclusions of the work.

\section{Polynomial Chaos Expansion}
\label{PCE section}
\subsection{Basic Aspects}
Assume a~probability space ($ \Omega, \pazocal{F}, \pazocal{P} $), where $ \Omega $ is an event space, $ \pazocal{F} $ is a~$ \sigma $-algebra on $ \Omega $ (collection of subsets closed under complementation and countable unions) and $ \pazocal{P} $ is a~probability measure on $ \pazocal{F} $.
If the input variable of a~mathematical model, $\pazocal{M}$, is a~random variable $ X(\omega) , \omega\in\Omega$,
the model response $Y$($ \omega $) is also a~random variable. Assuming that $ Y $ has a~finite variance, \PCE\ represents the output variable $ Y $ as a~function of an another random variable $ \xi $ called the \emph{germ} with given distribution
\begin{equation}
    Y=\pazocal{M}(X)=g^{\PCE}(\xi ),
\end{equation}
and representing the function $\pazocal{M}(X)$ via polynomial expansion in a~manner similar to the Fourier series of a~periodic signal. A~set of polynomials, orthogonal with respect to the distribution of the germ, are used as a~basis of the Hilbert space  $ L^2 $ ($ \Omega,  \pazocal{F}, \pazocal{P} $) of all real-valued random variables of finite variance, where  $ \pazocal{P} $ takes over the meaning of the probability distribution. The orthogonality condition for all  $ j \ne k $ is given by the inner product of $ L^2 $~($ \Omega,  \pazocal{F}, \pazocal{P} $) defined for any two functions $ \psi_j $ and $ \psi_k $ with respect to the weight function $ p_\xi $ (probability density function of $ \xi $) as:
\begin{equation}
    \langle	\psi_j,\psi_k\rangle
    =
    \int \psi_j(\xi)\psi_k(\xi)p_\xi  (\xi)
    \; \mathrm{d} \xi
    = 0.
\end{equation}

This means that there are specific orthogonal polynomials associated with the corresponding distribution of the germ via its weighting function.
For example, Hermite polynomials orthogonal to the Gaussian measure are associated with normally distributed germs.
Orthogonal polynomials corresponding to other distributions can be chosen according to Wiener-Askey scheme~\cite{Askey}. For further processing, it is beneficial to use normalized polynomials (orthonormal), where the inner product is equal to the Kronecker delta  $ \delta_{jk}$, i.e. $ \delta_{jk}=1$ if and only if $ j=k $, and  $ \delta_{jk}=0 $ otherwise
\begin{equation}
    \label{Eq: orthonormality}
    \langle	\psi_j,\psi_k\rangle=\delta_{jk}.
\end{equation}

In the case of $ \X $ and $\xxi$ being vectors containing $ M $ independent random variables, the polynomial $ \Psi (\xxi)$ is multivariate and it is built up as a tensor product of univariate orthogonal polynomials as
\begin{equation}
\label{Eq: MultVarPol}
    \Psi_{\balpha} ( \xxi )
    =
    \prod_{i=1}^{M}  \psi_{\alpha_i}(\xi_i),
\end{equation}
where $ {\balpha}\in \mathbb{N}^M $ is a~set of integers called the \emph{multi-index}. The quantity of interest (QoI), i.e., the response of the mathematical model $ Y=g(\X)$, can then be represented, according to Ghanem and Spanos \cite{Ghanem_spectral}, as
\begin{equation}
\label{PCE}
    Y = \pazocal{M}(\X) =
    \sum_{\balpha \in \mathbb{N}^M }
    \beta_{\balpha}\Psi_{\balpha}( \xxi),
\end{equation}
where $ \beta_{\balpha} $  are deterministic coefficients and $ \Psi_{\balpha} $ are multivariate orthogonal polynomials.

The main step in the solution procedure of determining the relation in Eq.~\eqref{PCE} is to determine the deterministic coefficients $\beta_\balpha$ to provide an accurate estimator.
In a practical context, an analyst usually only has access to input-output pairs that are generated by $\pazocal{M}$, rather than the full internal solver machinery (such as, e.g., mass or stiffness matrices)
Therefore, without losing generality, the rest of the text focuses on non-intrusive forms of \PCE. 
Nonetheless, note that the ensuing developments are equally applicable to the intrusive \PCE~formulations.
From a~statistical point of view, \PCE\ is a~simple linear regression model with intercept. Therefore, it is possible to use \emph{ordinary least square} (OLS) regression to minimize the error  $ \varepsilon $
\begin{equation}
    \bbeta
    =
    \argminA_{\bbeta \in \mathbb{R}^{P }}\, \frac{1}{\ns}\sum_{i=1}^{\ns}
    {
        \left[\bbeta^{T}
            \bPsi \left( \xxi^{(i)} \right)
            -  g  \left(  \x ^{(i)} \right)
        \right]
    }^2.
    \label{Eq: Chaos_OLS}
\end{equation}

Knowledge of vector $\bbeta$ fully characterizes the approximation via \PCE. To solve for $\bbeta$, first it is necessary to create $ \ns $ realizations of the input random vector $ \X $ and the corresponding results of the original mathematical model $ \pazocal Y  $, together called the experimental design (ED). Then, the vector of $P$ deterministic coefficients $\bbeta $ is calculated as
\begin{equation}
    \bbeta
    =
    (\bPsi^{T}\bPsi)^{-1} \ \bPsi^{T}  \pazocal Y ,
\label{Eq: PCe_OLS}
\end{equation}
where $ \bPsi $ is the data matrix
\begin{equation}
    \bPsi
    =
    \left\{
        \Psi_{ij}= \Psi_{j}(\xxi^{(i)}),  \;
        i=1, \ldots, \ns,   \;
        j=0, \ldots, P-1
    \right\}.
\label{Eq: basis_matrix}
\end{equation}

\subsection{Truncation of \PCE\ basis}
For practical computation, \PCE\ expressed in Eq.~\eqref{PCE} must be truncated to a~finite number of terms $ P $. Although it is generally possible to create a basis set using a tensor product of 1D polynomials, it leads to an extremely high number of basis functions.
This in its turn leads to a slow convergence of PCE construction. 
Therefore, the truncation is commonly achieved by retaining only terms whose total degree $ \vert \balpha \vert $ is less than, or equal to a~given $ p $. 
Therefore, the truncated set of \PCE\ terms using \emph{total polynomial order} is then defined as
\begin{equation}
    \pazocal A^{M,p}
    =
    \left\{
        {\balpha} \in \mathbb{N}^{M} : \left| {\balpha} \right|= \sum_{i=1}^{M} \alpha_i \leq p
    \right\}.
\label{Eq: truncation}
\end{equation}

The cardinality of the truncated \emph{index set} $ \pazocal A^{M,p} $ is given by
\begin{equation}
 \mathrm{card} \: \pazocal A^{M,p}= \frac{\left( M+p \right)!}{M! \: p!}.
\end{equation}

Moreover, in engineering applications, it is beneficial to prefer only basis functions with lower-order interaction terms.
This reduction of basis set is motivated by sparsity-of-effects principle, which states that a physical system is mostly affected only by main effects and low-order interactions. 
Therefore, it was proposed by Blatman and Sudret~\cite{BLATMANLARS} to create a~\PCE\ basis by a~\emph{hyperbolic} truncation scheme:
\begin{equation}
    \pazocal A^{M,p,q}=
    \left\{
        {\balpha}\in \mathbb{N}^{M} :
        || \balpha ||_q
        \equiv
        \Big( \sum_{i=1}^{M} \alpha_i^q \Big)^{1/q}
        \leq p
    \right\}.
    \label{Eq: Qtruncation}
\end{equation}

Using this truncation scheme can be graphically represented by selection of terms under the hyperbola parameterized by $q$. Note that $ q=1 $ corresponds to the standard truncation scheme according to Eq.~\eqref{Eq: truncation} and, for $ q<1 $, terms representing higher-order interactions are eliminated. Such an approach leads to a~dramatic reduction in the cardinality of the truncated set for high total polynomial orders $ p $ and high dimensions $ M $. 

When \PCE\ is truncated to a~finite number of terms, there is an error $ \varepsilon $ of the approximation such that
\begin{equation*}
    Y
    =
    \displaystyle \pazocal{M}{(\X)} =
    \sum_{\balpha \in \pazocal A}
    \beta_{\balpha} \Psi_{\balpha}(\xxi)
    +
    \varepsilon.
    \label{Eq: Chaos_Truncated_OLS}
\end{equation*}

\begin{figure}[h] 
	
	\centering
	\includegraphics[width=0.9\textwidth]{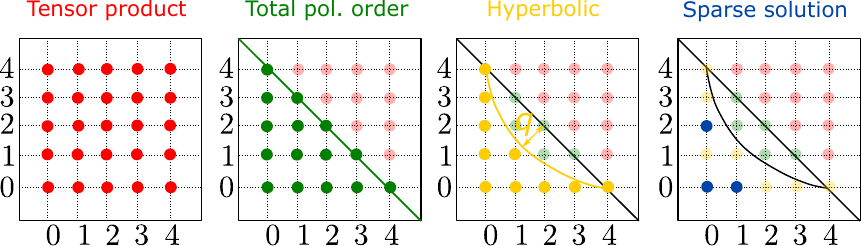}
	
	\caption{Comparison of various truncation schemes in 2D space ($M=2$ and $p=4$).}
	\label{Fig: Truncation PCE}
\end{figure}

Note that the number of terms $ P $ is highly dependent on the number of input random variables $ M $ and the maximum total degree of polynomials $ p $.
Estimation of $\bbeta$ by regression then needs at least the number of samples $\pazocal{O}( P \, \ln (P))$ for stable solution \cite{CohenOptimalWLS, NarayanOptimalWLS}.
Therefore, in case of a~large stochastic model, the problem can become computationally highly demanding.
However, one can utilize advanced model selection algorithms such as Least Angle Regression (LAR) \cite{LARS} to find an optimal set of \PCE\ terms and thus reduce the number of samples needed to compute the unknown coefficients if the true coefficient vector is sparse or compressible as proposed by Blatman and Sudret \cite{BLATMANLARS}.
Note that beside LAR, there are other best model selection algorithms such as orthogonal matching pursuit \cite{OMP} or Bayesian compressive sensing \cite{BCS} with comparable numerical results.
The \emph{sparse} set of basis functions obtained by any adaptive algorithm is further denoted for the sake of clarity as $ \A $.
From the comparison depicted in Fig. \ref{Fig: Truncation PCE}, one can see various sets of basis functions used for construction of PCE.
Tensor product and total polynomial order schemes are based only on the assumption of $p$, which might be even adaptive in the numerical algorithm \cite{Stahlbeton}.
Further, we can assume that general sparsity-of-effects principle is valid for the given model and use Hyperbolic truncation.
Finally, it is possible to construct sparse solution for the specific given model and the information matrix. 

\subsection{Estimation of Approximation Error}

Once the \PCE\ is constructed, it is crucial to estimate its accuracy.  A~common choice is the coefficient of determination $ R^2 $ (mean squared error normalized by the model variance), which is well known from machine learning. However for proper employment of $ R^2 $, it is necessary to create sufficiently rich validation set, which might be highly demanding from a computational viewpoint in industrial applications. In this paper, $ R^2 $ is used for comparison of the proposed methodology to a standard approach.

In the framework of an adaptive construction of \PCE\ , numerical algorithms are typically based on iterative construction of approximations using different hyper-parameters (e.g. $p$ and $q$) and the selection of the most suitable solution. The selection of the best approximation is based on the accuracy of \PCE\ , which can be used for the direct comparison among several \PCE{}s in order to choose the best surrogate model. Therefore it is beneficial to use methods which do not need any additional sampling of the original mathematical model. Unfortunately, $ R^2 $ may easily lead to overfitting in this case and thus advanced methods should be used. One of the most utilized methods for measuring the performance of the learning algorithm in recent years is the leave-one-out cross validation error $ Q^2 $. This statistic is based on residuals between the original surrogate model and the surrogate model built with the experimental design while excluding one realization.
This approach is repeated for all realizations in the experimental design and the average error is estimated. Although the calculation of $ Q^2 $ is typically highly time-consuming, it is possible to obtain results analytically from a~single \PCE\ \cite{BLATMAN2010}.

\subsection{Post-processing of \PCE}
\label{moments}
The specific form of \PCE\ together with the orthogonality of the polynomials allows for a~powerful and efficient post-processing. Once a~\PCE\ approximation is created, it is possible to analytically obtain statistical moments or sensitivity indices of the QoI. Generally, a statistical moment of $m$th order is defined as:

\begin{align*}
\big<Y^{m} \big> = \int  \big[g \big(\mathbf{X}\big) \big]^{m}p_{ \mathbf{X} } \big( \mathbf{x} \big)d\mathbf{x}= 
\int  \big[ \sum_{\alpha\in  \mathbb{N}^{ M}} {{\beta_{\alpha}}}{ {\Psi_{\alpha}(\mathbf{\xi})}} \big]^{m}p_{ \xi } \big( \xi \big)d\xi=\\
=\int \sum_{\alpha_{1} \in  \mathbb{N}^{ M}}...\sum_{\alpha_{m} \in  \mathbb{N}^{ M}} \beta_{\alpha_{1}}...\beta_{\alpha_{m}}\Psi_{\alpha_1}(\mathbf{\xi})...\Psi_{\alpha_m}(\mathbf{\xi})p_{ \xi } \big( \xi \big)d\xi=\\
=\sum_{\alpha_{1} \in  \mathbb{N}^{ M}}...\sum_{\alpha_{m} \in  \mathbb{N}^{ M}} \beta_{\alpha_{1}}...\beta_{\alpha_{m}} {{\int \Psi_{\alpha_1}(\mathbf{\xi})...\Psi_{\alpha_m}(\mathbf{\xi})p_{ \xi } \big( \xi \big)d\xi }}.
\end{align*}

As can be seen from the final part of the formula, in case of PCE, it is necessary to integrate over basis functions (orthonormal polynomials), which leads to dramatic simplification in comparison to the integration of the original mathematical function. Moreover, it is well known that \PCE{} allows for analytical solution of the associated integral. Besides well known formulas for mean ($\mu_Y=\beta_0$) and variance ($\sigma_Y^2=\sum_{{\balpha} \in {\A}} {\beta_{\balpha}^2} - \beta_0^2$), higher statistical central moments
    skewness $\gamma_Y$ ($ 3^{\text{rd}}$ moment) and
    kurtosis $\kappa_Y$ ($ 4^{\text{th}}$ moment)
 can be also obtained using analytical formulas for Legendre and Hermite polynomials \cite{NOVAK2022106808}.  Note that \PCE{} is in the identical form as Hoeffding-Sobol decomposition of a function and thus it is possible to easily derive also conditional variances of any order and corresponding Sobol indices \cite{Sobol:76, SUDRET}.

 Finally, the \PCE{} approximation can be also exploited for an estimation of a probability distribution of QoI. A first possible approach to build the PDF of QoI consists in directly evaluating a \PCE{} for a large number of samples of input random vector and processing of the corresponding results by kernel density estimation (KDE) \cite{KDERosenblueth}. Although the combination of \PCE{} and KDE is often utilized for UQ \cite{TORRE2019601, WANG2021113854}, it might be complicated to selected appropriate kernel function and band-width hyper-parameter leading to accurate identification of PDF.

 A second general approach, further extended in this study, is based on approximations of PDF/CDF by analytical functions parameterized by statistical moments derived directly from \PCE. A simple approximation can be in form of Gram-Charlier (G-C) expansion or similar Edgeworth series containing one more Hermite polynomial than G-C, both based on the first four statistical moments \cite{GramConditions}. Similarly as in case of Sobol indices, conditional distributions can be easily obtained from PCE \cite{wan2006multi} as well as advanced distribution-based sensitivity indices \cite{CANIOU20107625,NOVAK2022106808}. The second approach has also two main drawbacks: analytical formulas are known only for some polynomials and it is computationally efficient to estimate only the first four statistical moments. The later significantly limits utilization of the advanced analytical PDF approximations typically based on higher number of statistical moments.

\section{Fractional Moments from Polynomial Chaos Expansion}~\label{sec:FractionalMoments}
In this section, we introduce our proposed method to determine the probability density function $f_Y(y)$ of a random variable $Y$ based on the post-processing of the PCE in the form of fractional moments. Recall in this context that the $r$-th absolute fractional moment of the random variable $Y$ is defined as:
\begin{equation}
\mathbb{E} \left[ |Y|^r \right]= \int_{-\infty}^\infty |y|^r \ f_{Y}(y) \ dy,
\label{eq: fracmoment}
\end{equation}
where $r$ can be any real number.
Clearly, when $r$ in Eq.~\ref{eq: fracmoment} takes an integer value, the equation reduces again to the description of a general moment, making Eq.~\ref{eq: fracmoment} in essence a generalization of the well-known concept of statistical moments.

The main advantage of working with fractional moments, is that $\mathbb{E} \left[ |Y|^r \right]$ carries information about an infinite number of discrete moments. 
This can be understood by first performing a Taylor series expansion of $|Y|^r$ around its mean value $\mu_Y = E[|Y|]$:
\begin{equation}
    |Y|^r = \sum_{i=0}^\infty \binom{r}{i}\mu_y^{(r-i)}\left(y - \mu_Y \right)^i,
    \label{eq: Taylor_Fractional_Moment}
\end{equation}
with $i$ any non-negative integer, $\mu_y^{(r-i)}$ the expected value of $|Y|^{(r-i)}$ and the fractional binomial $\binom{r}{i}$ given by:
\begin{equation}
\binom{r}{i} = \frac{r(r-1)(r-2)\ldots(r-i+1)}{i(i-1)(i-2)\ldots1}.  
\end{equation}

Taking the expectation of both sides of Eq.~\ref{eq: Taylor_Fractional_Moment} yields:
\begin{equation}
        \mathbb{E}\left[|Y|^r\right] = \sum_{i=0}^\infty \binom{r}{i}\mu_y^{(r-i)}\mathbb{E}\left[\left(y - \mu_Y \right)^i\right],
\end{equation}
from which can be seen that the right-hand side indeed contains an infinite number $i=1,\ldots,\infty$ of integer moments while the left-hand side of the equation is the $r^{th}$ fractional moment of $Y$.
In this equation, the term $\binom{r}{i}\mu_y^{(r-i)}$ can be thought of as a sort of weight that is assigned to the integer moment in the series expansion that describes the fractional moment. 
In this context, observe that when $i$ is fixed, $\binom{r}{i}\mu_y^{(r-i)}$ increases as $r$ increases, whereas when $r$ is fixed, the value of $\binom{r}{i}\mu_y^{(r-i)}$ decreases when $i$ increases.
This indicates that the higher the fractional order $r$,
the greater the contribution of higher-order integer moments to the $r^{th}$ fractional moment value.
To effectively estimate $f_Y(y)$ from the PCE, it is as such important to estimate higher-order fractional moments.
At the same time, it is important to keep in mind that these higher-order fractional moments are much more difficult to obtain than lower-order fractional moments.
This trade-off needs to be addressed case-by-case when applying the proposed technique.

\subsection{Estimation of the Fractional Moments via H\"{o}lder's inequality}

Direct numerical estimation of fractional moments by Monte Carlo approach could be computationally expensive, especially in engineering applications. 
However, the estimation can be significantly accelerated by approximation in form of H\"{o}lder's inequality:

\begin{equation}
    \mathbb {E} {\bigl [}|Y|^{r}{\bigr ]}\leqslant \left(\mathbb {E} {\bigl [}|Y|^{s}{\bigr ]}\right)^{\frac {r}{s}}.
    \label{Eq. HoldersInequality}
\end{equation}
H\"{o}lder's inequality is often utilized for estimation of error bounds in various applications of theory of probability, however it can be also utilized for an efficient estimation of fractional moments from standard integer statistical moments. 

However, it might be still computationally expensive to estimate higher integer moments by crude Monte Carlo methods. Therefore we propose to combine H\"{o}lder's inequality with \PCE{} surrogate model, as the latter is well-known for an accurate and an efficient statistical analysis of QoI. Such approach should be significantly more stable in comparison to sampling methods, since the first four statistical moments can be obtained analytically from \PCE{} coefficients, i.e. $s\in\left[1,2,3,4 \right]$. It is clear that the error of the approximation grows with the difference $\rvert s-r \lvert$. Therefore, an integer moment $\mathbb {E} {\bigl [}|Y|^{s} \bigr]$ utilized for the estimation of a fractional moment should be selected as close as possible to the selected $r$. Note that H\"{o}lder's inequality according to Eq.~\eqref{Eq. HoldersInequality} is valid only for $1<r<s<\infty$ and thus, for the sake of clarity, it is necessary to use Eq. \eqref{Eq. HoldersInequalityReverse}, if the nearest integer moment $s>r$.

\begin{equation}
    \mathbb {E} {\bigl [}|Y|^{s}{\bigr ]}\geqslant \left(\mathbb {E} {\bigl [}|Y|^{r}{\bigr ]}\right)^{\frac {s}{r}}.
    \label{Eq. HoldersInequalityReverse}
\end{equation}

Naturally it is possible to reliably estimate fractional moments only in the interval between integer moments obtained from \PCE{}, i.e.  $r\in(1,4)$. Estimated fractional moments can be further used for various tasks including statistical or sensitivity analysis similarly as integer moments \cite{NOVAK2022106808,SUDRET}. 

\subsection{Description of the PDF based on the fractional moments}
Fractional moments are especially important for estimation of the most suitable probability distribution of the QoI. Although it might be sufficient to fit a selected well-known distribution in simple applications, artificial distributions parametrized by statistical moments are more flexible and can capture more complicated shapes of probability distributions. 
Some of the simplest parametrized distributions are the Gram-Charlier expansion or Edgeworh series based on perturbation of Gaussian distribution exploiting information from the first four statistical moments \cite{GramConditions}. It was shown, that Gram-Charlier expansion is efficient especially in combination with PCE, since we can obtain necessary statistical moments analytically \cite{NOVAK2022106808}. However, once the fractional moments are estimated directly from PCE, it is possible to use more advanced and flexible distribution models such as recently proposed mixture of extended inverse Gaussian and log extended skew-normal distributions (M-EIGD-LESND) \cite{DING2023109775}, which is described as:
\begin{multline}
 f_{M-EIGD-LESND}(x;\vartheta)= \\
 w\eta\sqrt{\frac{b}{2\pi}}x^{-\eta/2-1}\exp{\left[-\frac{b(x^\eta-a)^2}{2x^\eta a^2}\right]}+
(1-w)\frac{1}{dx}\phi\left(\frac{log(x)-c}{d}\right)\frac{\Phi(\tau\sqrt{1+\theta^2}+\theta\frac{log(x)-c}{d})}{\Phi(\tau)},\\
 ~~\text{with}~x>0,
\end{multline}
where $\phi(\cdot)$ and $\Phi(\cdot)$ are the probability density and cumulative density functions associated with a standard Gaussian distribution, respectively, and $log(\cdot)$ denotes natural logarithm.
The M-EIGD-LESND contains a set of 8 free parameters, $\left\lbrace w, \eta, a, b, c, d, \theta, \tau \right\rbrace$, which are obtained by a matching of estimated fractional moments and fractional moments of M-EIGD-LESND. 
This task requires to solve system of non-linear equations by any numerical solver (see \cite{DING2023109775} for more details). 
It is noteworthy that the $r^{th}$ fractional moment of the M-EIGD-LESND function, $M^r_{X_{M-EIGD-LESND}}$ can be analytically determined as~\cite{DING2023109775}:
\begin{multline}
    M^r_{X_{M-EIGD-LESND}} = w \exp{\left[\frac{b}{a}\right]}\sqrt{\frac{2b}{\pi}}a^{r/\eta-0.5}K_{0.5-r/\eta}\left(\frac{b}{a}\right)+(1-w)\exp{\left( cr+0.5d^2r^2\right)}\frac{\Phi\left(\tau + \frac{\theta dr}{\sqrt{1+\theta^2}}\right)}{\Phi(\tau)},
\end{multline}
with $K_\alpha(\beta)$ the modified Bessel function of the second kind.
The thus approximated distribution function can then be used further for reliability analysis or distribution-based sensitivity analysis. 
The M-EIGD-LESND is adopted in numerical examples to measure the error of the proposed method and traditional Monte Carlo approach.

\subsection{Numerical Algorithm}
The proposed approach allows for significant extension of a statistical or sensitivity analysis of costly mathematical models. Specifically, estimated fractional moments can be used for an approximation of probability distribution of QoI as summarized in the following pseudo-algorithm employed in the numerical examples.
In the algorithm, the vector $\mathbf{r}$ contains all fractional moments considered in the analysis (whose dimensionality is 8) and $\lceil \cdot \rfloor$ denotes rounding to the closest integer.

\begin{algorithm}[!ht]
	\caption{Estimation of fractional moments by PCE and construction of a probability distribution}
\begin{algorithmic}[1]
\Statex	\textbf{Input:}   experimental design (ED) with samples of $\X$ and $\pazocal Y=\pazocal{M}(\X)$, set of basis functions $\A$

\State get ${\bbeta }$ by OLS

\State get $\mu_Y$, $\sigma^2_Y$, $\gamma_Y$ and $\kappa_Y$ analytically from ${\bbeta }$
\For{$r$ in $\mathbf{r}$}
\State find a nearest integer moment $s=\lceil r \rfloor$
\State get approximated $r$th fractional moment by Eq.~\eqref{Eq. HoldersInequality} (or Eq. \eqref{Eq. HoldersInequalityReverse} if $r>s$)
\EndFor
\State get parameters of M-EIGD-LESND from fractional moments \cite{DING2023109775}
\Statex	\textbf{Output:} $\mathbb {E} {\bigl [}|Y|^{\mathbf{r}}{\bigr ]}$ and corresponding PDF/CDF of M-EIGD-LESND
                        
\end{algorithmic}
\label{Alg:Fractional moments}
\end{algorithm}

\clearpage
\section{Numerical Examples}~\label{sec:NumericalExamples}
The proposed approach is presented in three numerical examples of increasing complexity and which illustrated different aspects of the approach. The proposed approach is utilized for estimation of the following fractional moments $ \mathbb{E}\left[|Y|^r\right], \, r \in \mathbf{r}=\left[1.1,1.2,1.8,1.9,2.1,2.2,2.9,3\right]$. Note that the fractional moments are close to the integer moments obtained analytically from \PCE{} in order to reduce the error of approximation by H\"{o}lder's inequality. Fractional moments are further used for identification of the most suitable probability distribution as described in the previous section. The \PCE\ is constructed using the \textsf{UQPy} package \cite{TSAPETIS2023101561}. The obtained results of the proposed approach are compared to approximation based on standard sampling approach represented by Latin Hypercube Sampling (LHS) \cite{Conover:LHS:75,Owen:CLT:LHS:92}. Naturally, one can use various advanced or adaptive sampling schemes \cite{SHIELDS201696,AdaptiveCoherence,NOVAK2021114105} instead of LHS for achieving higher accuracy, however this task is beyond the scope of this paper. Both approximations are constructed for increasing number of simulations.
To compare both methods, we estimate error by non-negative Kullback-Leibler divergence $D_{\mathrm{KL}}\left(Y || \tilde{Y} \right)$ of a reference CDF $ F_Y$ and an approximated CDF $F_{\tilde{Y}}$. The error is calculated on CDF for improved numerical stability \cite{PARK20122025} as implemented in SciPy \cite{Scipy}:
\begin{equation}
D_{\mathrm{KL}}\ \left(Y || \tilde{Y} \right)=  F_Y(\chi) \ln \frac{F_{Y}(\chi)}{F_{\tilde{Y}}(\chi)}  + F_{\tilde{Y}}(\chi) -F_Y(\chi),
\end{equation}
where $\chi\in\mathbb{R}$. The total error $\epsilon$ is than obtained by integration of $D_{\mathrm{KL}}$ simply as:

\begin{equation}
    \epsilon = \int_{\mathbb{R}} D_{\mathrm{KL}}\ \left(Y || \tilde{Y} \right)\, \der{\chi}
\end{equation}
In order to get reliable statistical information on convergence, we run $n_{\mathrm{stat}}=100$ repetitions of the algorithm and plot $\mathbb{E} [\epsilon]\pm{\sigma}$ interval of the obtained errors.

\subsection{Academic Example: Gaussian Distribution}

The very first example shows the convergence of the proposed method. This toy example is represented by a simple analytical function of input random vector containing three independent Gaussian variables $\mathbf{X}\sim \pazocal{N}(\boldsymbol{\mu}=10, \boldsymbol{\sigma^2}=4)$ and thus the quantity of interest is also a Gaussian variable $Y\sim \pazocal N(\mu=50, \sigma^2=12)$:

\begin{equation}
    Y=20+X_1+X_2+X_3.
\end{equation}

\begin{figure}[!h]
    \centering
    \includegraphics[width=1\textwidth]{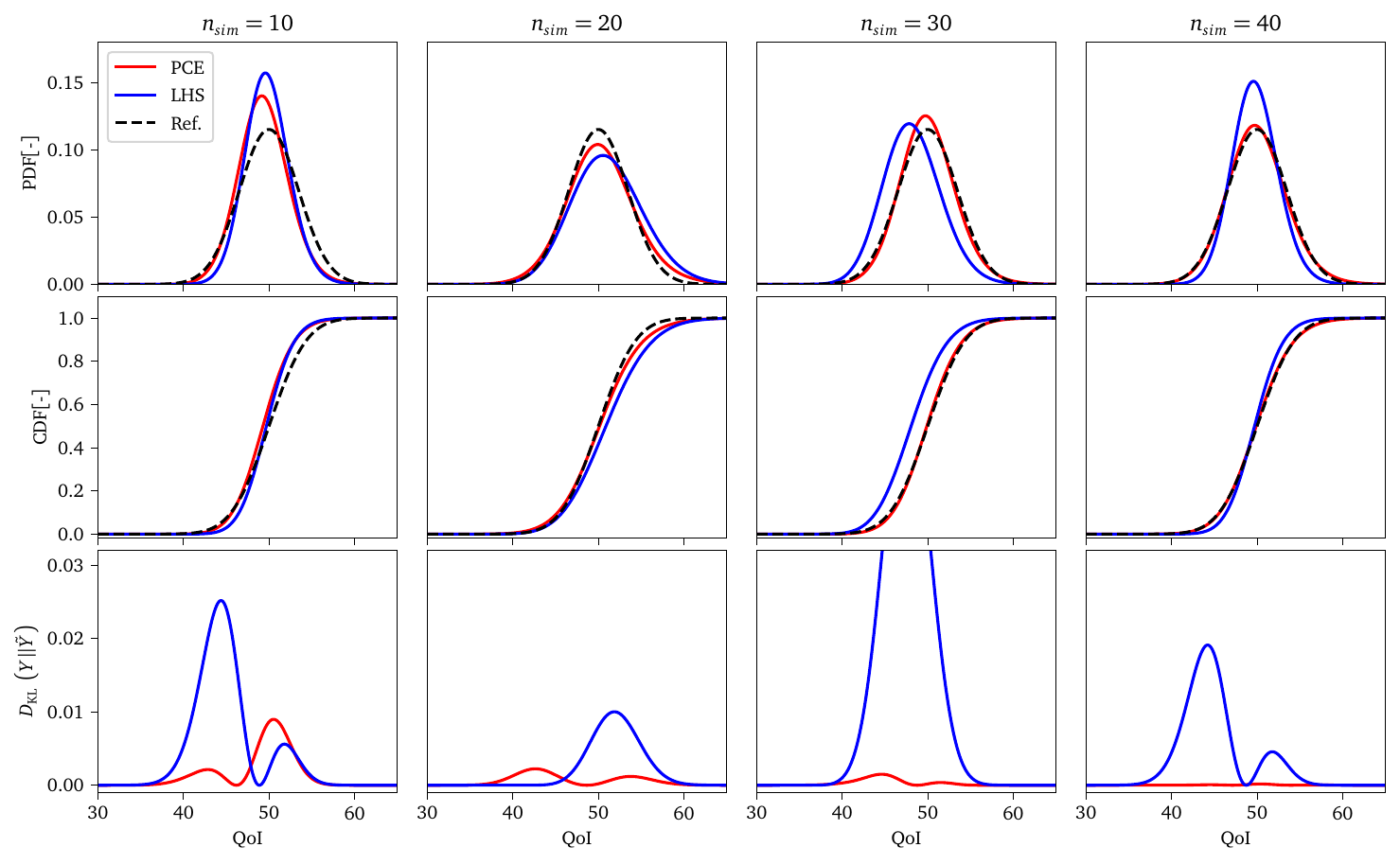}
    \caption{Typical realization of results for the first example (Gaussian distribution). The rows show estimated PDFs, CDFs and errors in approximations respectively. Each column corresponds to increasing number of simulations used for estimation of fractional moments and probability distributions.}
    \label{fig: Gauss_realization}
\end{figure}

Typical realizations of identified probability distributions based on fractional moments estimated by  LHS and the proposed approach for incresing size of ED can be seen in Fig. \ref{fig: Gauss_realization}. Since it is possible to analytically obtain reference distribution, there is not any error caused by approximation of the probability  distribution.  Although the resulting distribution of QoI is very simple, the estimated distributions converge slower then expected due to over-parametrized approximating function taking 8 fractional moments into account. Utilized approximation is universal parametrized distribution suitable for various types of distribution, nonetheless it might be over sophisticated for approximation of a simple distribution and a simple models should be preferred, e.g. the well-known Gram-Charlier expansion \cite{GramConditions}. The convergence of LHS is further affected by sensitivity to outliers, while the proposed method via \PCE{} is clearly more stable. General convergence of both methods can be seen in Fig. \ref{fig: Convergence_plots} a). Note that besides higher accuracy in mean values, the variance of the proposed method is significantly smaller in comparison to LHS for $n_{\mathrm{sim}}>35$ samples.

\begin{figure}[!h]
    \centering
    \includegraphics[width=1\textwidth]{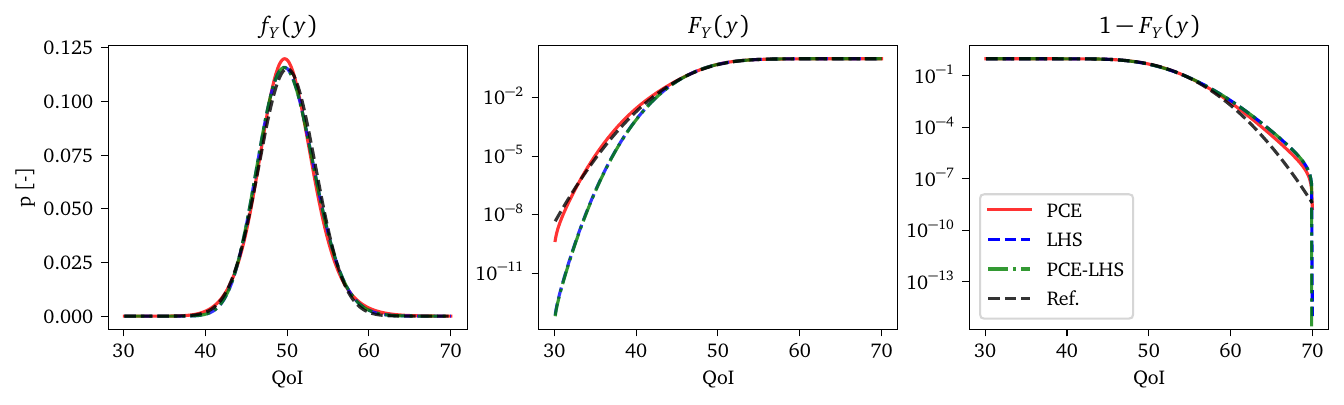}
    \caption{Convergence study for $n_{\mathrm{sim}}=200$. }
    \label{fig: Gauss_tails}
\end{figure}

The stability of the proposed method can be also seen in Fig. \ref{fig: Gauss_tails} showing obtained distributions for $n_{\mathrm{sim}}=200$ simulations. Although \PCE{} leads to slight error near mean value, it leads to almost perfect accuracy at both tails of the CDF. Although LHS is very efficient method for estimation of mean values, it has clearly worse performance in estimation of higher moments affected by tails and thus also fractional moments. Note that although \PCE{} is based on identical samples, it is an approximation of QoI over whole input space and thus its result is less affected by outliers. This fact is also supported by results obtained from LHS sampling using \PCE{} surrogate model instead of the original mathematical model (PCE-LHS). The results of PCE-LHS are identical to LHS with original model, which clearly shows that although the surrogate model is accurate, LHS sampling adds additional error to estimated fractional moments and thus it is beneficial to the proposed approach instead of numerical estimation if a \PCE{} is available.  

\subsection{Finite Element model of a Plate}
The second case study deals with a model of a thin steel plate of $1$ [m] by $1$ [m] that is fully clamped at one side.
The plate is subjected to a distributed load over the top surface, and its displacement $\bs{u}$ is computed using a finite element model consisting of 100 evenly distributed linear shell elements, resulting in 121 nodes.
As such, there are 110 active nodes in the model.
In the analysis, the degrees of freedom per node correspond to one translation and two rotations.

The corresponding equilibrium equation associated with the finite element model of the plate is represented as:
\begin{equation}
    \bs{K}(\bs{\theta})\bs{u} =\bs{f},
\end{equation}
with $\bs{K}\in\mathbb{R}^{330 \times 330}$ the stiffness matrix of the plate; $\bs{\theta} = [E, t]$, with $E$ representing Young's modulus and $t$ the thickness of the plate; $\nu$ Poisson's ratio, $\bs{f}\in\mathbb{R}^{330}$ is a vector collecting the forces acting on the nodes of the FE model; and $\bs{u}\in\mathbb{R}^{330}$ the resulting displacement vector. We assume $E, t, \nu$ to be (truncated) Gaussian variables with vector of mean values $\boldsymbol{\mu}^T=[\num{2.1e11}, \num{5e-3}, 0.3]$ with vector of coefficients of variation $\mathbf{v}=[0.15,0.1,0.1]$. It is assumed that the degrees-of-freedom of the finite element model have been ordered such that the first 110 components of $\bs{u}$ correspond to vertical displacements.

\begin{figure}[!h]
    \centering
    \includegraphics[width=1\textwidth]{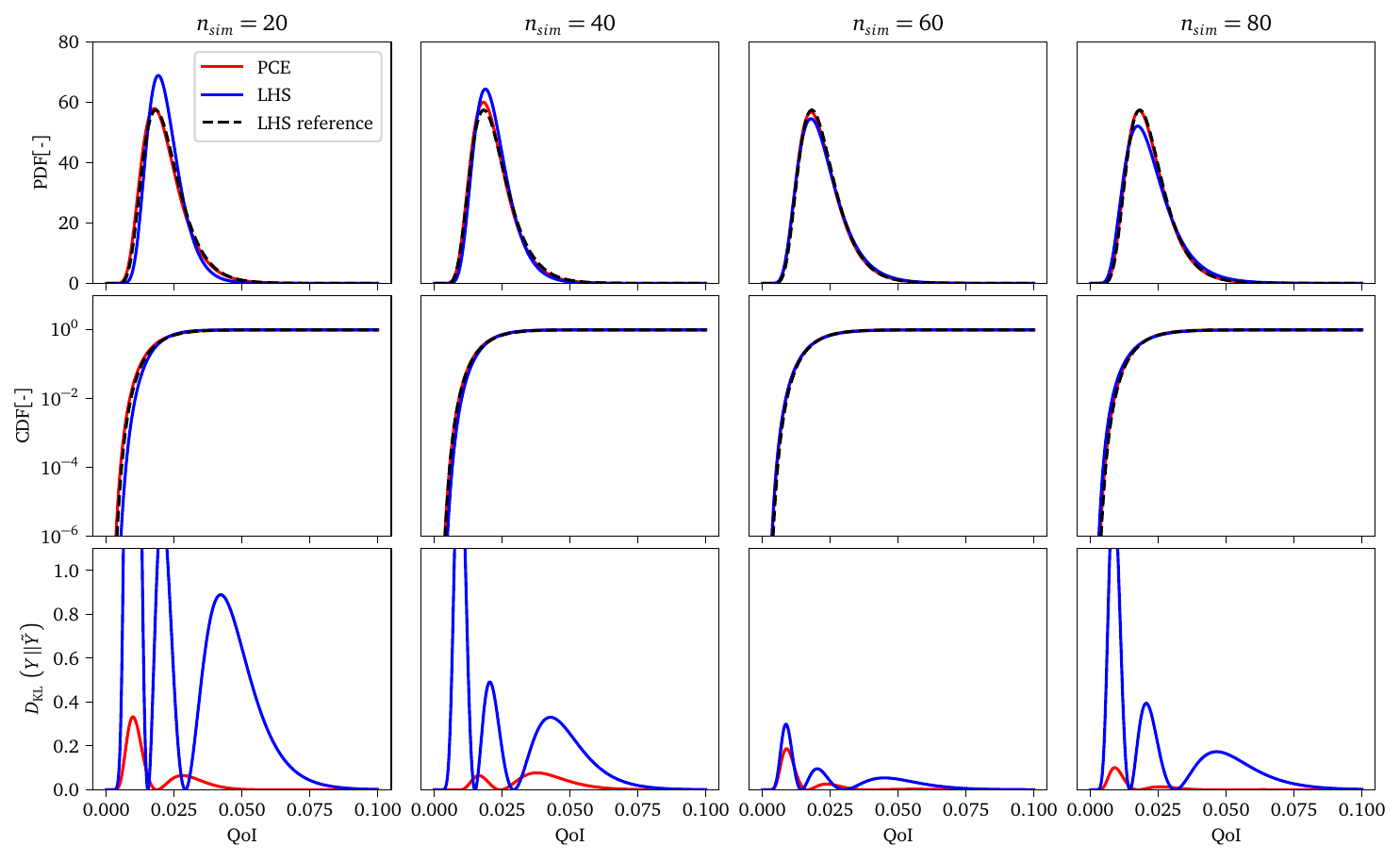}
    \caption{Typical realization of results for the second example (FEM of a plate). The rows show estimated PDFs, CDFs and errors in approximations respectively. Each column corresponds to increasing number of simulations used for estimation of fractional moments and probability distributions.}
    \label{fig: plate_comparison}
\end{figure}

Selected realizations of approximations are compared in Fig. \ref{fig: plate_comparison}. In contrast to the previous example, a reference solution cannot be obtained analytically and thus we use empirical CDF obtained by LHS with $n_{\mathrm{sim}}=10^6$ simulations. The proposed approach leads to very accurate approximation of target distribution already for $n_{\mathrm{sim}}=20$ samples in contrast to standard LHS. Moreover the convergence of the accuracy for increasing $n_{\mathrm{sim}}$ is stable, while standard LHS approach achieves lower accuracy for $n_{\mathrm{sim}}=80$ in comparison to $n_{\mathrm{sim}}=60$. Identical behavior of both methods can be seen also in statistical results compared in Fig.\ref{fig: Convergence_plots} b). The proposed approach achieves both lower mean and variance of the estimated total error.

\subsection{Dynamic Car Model}
The third case study in this paper considers a so-called quarter-car model.
This is a 2 degree of freedom idealisation of the dynamics of the suspension of a moving car. 
Specifically, this case study is concerned with assessing the distribution of the comfort of the vehicle suspension, given the uncertainty in some of the properties of the system.
The quarter-car dynamics can be represented as a set of two ordinary differential equations:
\begin{align}
    m_s\ddot{x_s} + c_s (\dot{x}_s - \dot{x}_{us}) + k_s(x_s - x_{us}) = 0,\\
    m_{us}\ddot{x}_{us} - c_s (\dot{x}_s - \dot{x}_{us}) -  k_s(x_s - x_{us})+ c_t(\dot{x}_{us}-\dot{x}_0) + k_t(x_{us}-x_0) = 0,
\end{align}
with $\dot{\bullet}$ denoting the time derivative of $\bullet$, $x_{us}$ the displacement of the unsprung mass (i.e., the suspension components, wheel and other components directly connected to them); $x_s$ the displacement of the sprung mass (i.e., all components resting on the suspension); $m_{us}$ and $m_s$ the unsprung and sprung mass of a quarter of the car; $c_s$ and $c_t$ respectively the damping coefficients of the suspension and tire; $k_s$ and $k_t$ respectively the stiffness coefficients of the suspension and tire.
We assume $c_s, k_s, k_t$ to be (truncated) Gaussian variables with vector of mean values $\boldsymbol{\mu}^T=[\num{1e4}, \num{4.8e4}, \num{2e5}]$ with identical coefficient of variation $v=10\%$.
Finally, $x_0$ and $\dot{x}_0$ are the displacement and velocity in vertical direction that excite the bottom of the wheel (i.e., the road profile).
The complete road profile is denoted by $x_0(t)$, with $t$ denoting the simulation time.

To solve this coupled system of ODEs, a state-space model is employed:
\begin{equation}
        \frac{d}{dt} \left[\begin{matrix}
            x_{us}- x_0\\
            \dot{x}_{us}\\
            x_{s}-x_{us}\\
            \dot{x}_s\\
        \end{matrix}\right]
        = \bs{A}
        \left[\begin{matrix}
            x_{us}-x_0\\
            \dot{x}_{us}\\
            x_s-x_{us}\\
            \dot{x}_s
        \end{matrix}\right] + 
        \left[\begin{matrix}
            -1\\
            \frac{4c_t}{m_{us}}\\
            0\\
            0
        \end{matrix}\right]
        \dot{x}_0,
\end{equation}
with  the matrix $\bs{A}$ equal to:
\begin{equation}
    \bs{A} = \left[\begin{matrix}
        0 & 1 & 0 & 0\\
        \frac{-4k_t}{m_{us}} & \frac{-4(c_s+c_t)}{m_{us}} & \frac{4k_s}{m_{us}} & \frac{4c_s}{m_{us}}\\
        0 & -1 & 0 & 1\\
        0 & \frac{4c_s}{m_s} & \frac{-4k_s}{m_s} & \frac{-4c_s}{m_s}
    \end{matrix}\right].
\end{equation}

Four state variables are considered, being respectively the tire deflection ($ x_{us}- x_0$); the unsprung mass velocity $ \dot{x}_{us}$; the suspension stroke $ x_{s}-x_{us}$, and sprung mass velocity $\dot{x}_s$.
Typically, in the context of assessing the dynamical comfort of a car, two parameters are of interest: the suspension stroke (i.e., the relative displacement of the car body with respect to the unsprung mass) and the acceleration of the sprung mass. In the proceeding study, the damping effect of the tire, $c_t$ is considered negligible. 
The limit state function in this example is based on the first excursion event of the suspension stroke, and explicitly defined as:
\begin{equation}
    1-\max_t\left(\frac{1}{x_c} \left\lvert x_s(t)-x_{us}(t) \right\rvert\right),
\end{equation}
with $x_c=30~mm$ the threshold value for the stroke.

\begin{figure}[!h]
    \centering
    \includegraphics[width=1\textwidth]{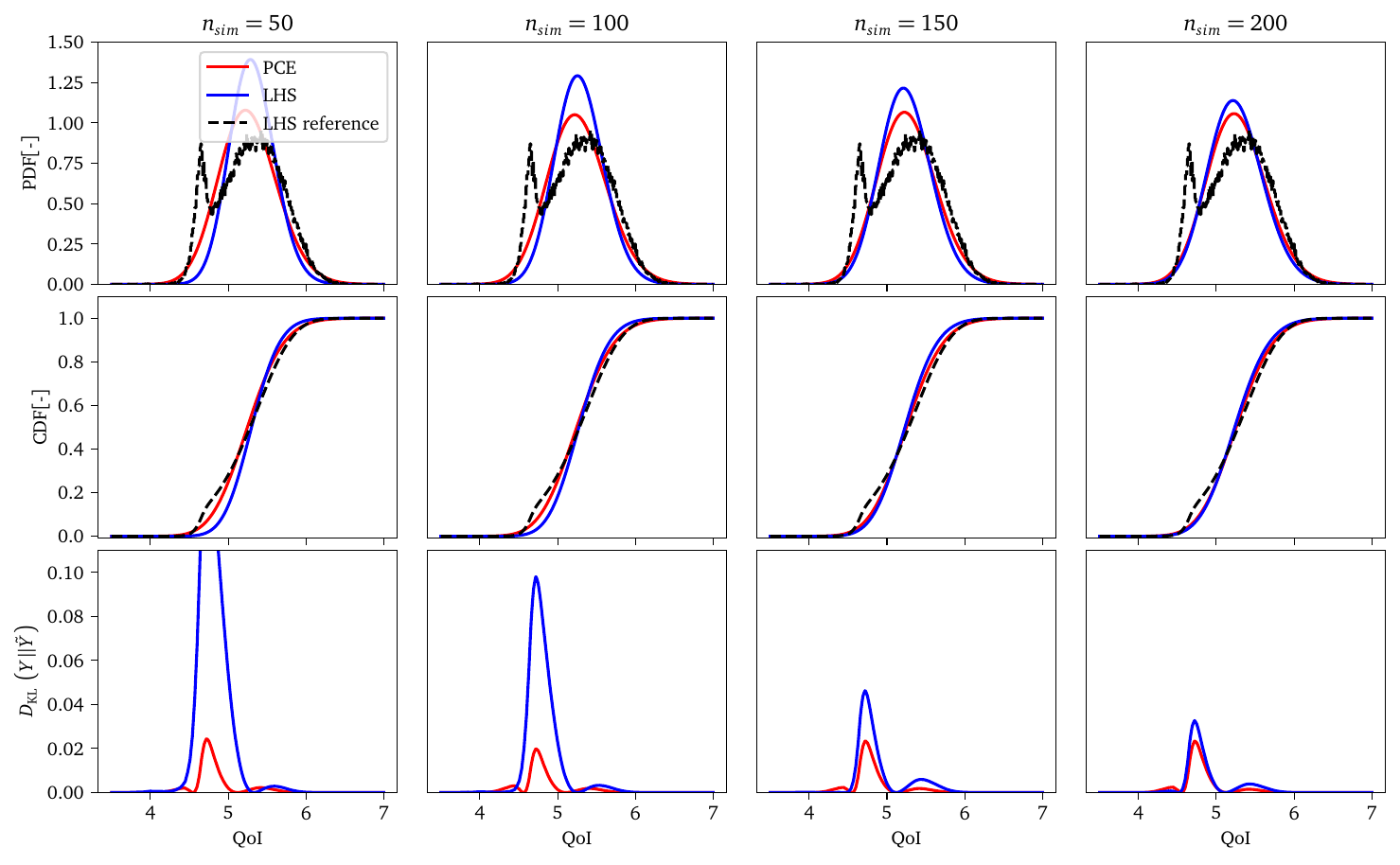}
    \caption{Typical realization of results for the third example (dynamic car model). The rows show estimated PDFs, CDFs and errors in approximations respectively. Each column corresponds to increasing number of simulations used for estimation of fractional moments and probability distributions.}
    \label{fig: car_comparison}
\end{figure}

\begin{figure}[!h]
    \centering
    \includegraphics[width=1\textwidth]{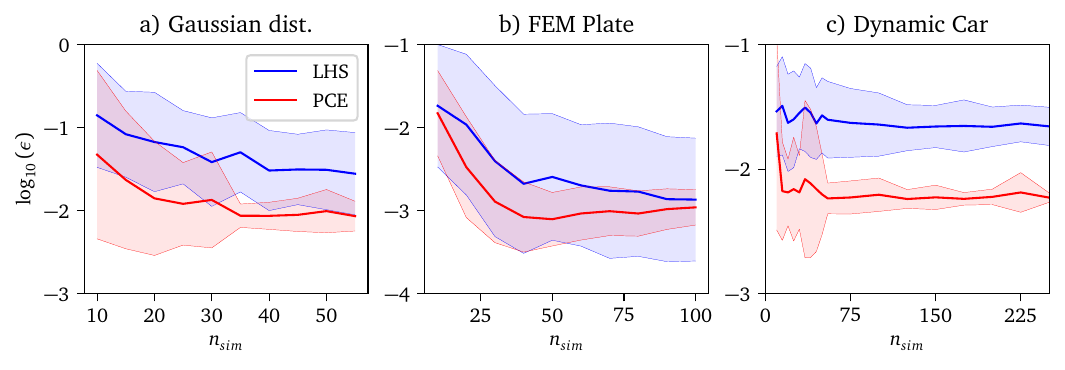}
    \caption{Convergence plots showing mean values and $\pm\sigma$ interval of the total error $\epsilon$ obtained by the proposed method and standard LHS approach for a) Gaussian model, b) plate model and c) Dynamic car model.}
    \label{fig: Convergence_plots}
\end{figure}

Similarly as in the previous examples, selected realizations of approximations for increasing $n_{\mathrm{sim}}$ can be seen in Fig. \ref{fig: car_comparison}. Note that the distribution of QoI is bimodal (as can be clearly seen from empirical CDF) and thus it can not be accurately approximated by the adopted M-EIGD-LESND. However, the obtained results show very fast convergence of the proposed approach to the optimal solution with minimum possible error, while standard LHS converges to the optimum significantly slower as can be seen in the last column showing the obtained distributions for $n_{\mathrm{sim}}=200$. Statistical results compared in Fig. \ref{fig: Convergence_plots} show stable accuracy of the proposed method for $n_{\mathrm{sim}}>60$, while LHS significantly larger variance and lower mean accuracy. Although this general behavior of standard LHS could be seen also in the previous example, it is amplified in the last example by the fact that fractional moments of the investigated bimodal distribution are even more affected by position of samples in the input random space.

\section{Conclusions \& Further Work}~\label{sec:Conclusions}
A novel approach for estimation of fractional moments directly from polynomial chaos expansion was proposed in this paper. The proposed method combines well-known formulas for estimation of integer statistical moments from PCE coefficients together with H\"{o}lder's inequality in order to analytically obtain arbitrary fractional moments. The fractional moments were further used for a construction of probability distribution based on adopted M-EIGD-LESND algorithm.  Obtained results from the presented numerical examples clearly show that the proposed method leads to stable and accurate estimations. Moreover, it achieves also a superior computational efficiency in comparison to a standard method based on Latin hypercube sampling. Therefore, it can be concluded that an error caused by an H\"{o}lder's inequality approximation is typically lower than error caused by discrete sampling methods, at least for low-size ED. Naturally, the benefits of the proposed method will be crucial for distribution-based sensitivity measures typically based on differences between conditional probability distributions, which can be accessed from a single PCE \cite{NOVAK2022106808}. On the other hand, it is well-known that PCE suffers from \emph{curse of dimensionality} and thus the proposed approach is not suitable for high-dimensional applications. There are also still some important topics for further research. First of all, the accuracy of PCE is highly dependent on the type of sampling scheme \cite{LuthenReview} and thus it will be necessary to investigate the most suitable sampling schemes and/or active learning algorithms with respect to an estimation of statistical moments \cite{NOVAK2021114105}.

\section*{Acknowledgments}
The  first author  acknowledges  financial  support  provided  by  by the Czech Science Foundation under project number 22-00774S. 

\bibliographystyle{elsarticle-num}
\bibliography{literatura}

\end{document}